\documentclass[12pt]{article}
\usepackage{scicite}
\usepackage{times}
\usepackage{amsmath,amssymb}
\usepackage{graphicx}
\usepackage{color}
\usepackage[normalem]{ulem}	
\newenvironment{sciabstract}{%
\begin{quote} \bf}
{\end{quote}}

\newcounter{lastnote}

\newcommand{\p}{\partial}

\begin{document}

\author{Gary P. T. Choi$^{1}$, Levi H. Dudte$^{1}$, L. Mahadevan$^{1,2,3\ast}$\\
\\
\footnotesize{$^{1}$John A. Paulson School of Engineering and Applied Sciences, Harvard University}\\
\footnotesize{$^{2}$Departments of Physics, and Organismic and Evolutionary Biology, Harvard University}\\
\footnotesize{$^3$ Kavli Institute for Nanobio Science and Technology, Harvard University, Cambridge, MA 02138, USA}\\
\footnotesize{$^\ast$To whom correspondence should be addressed; E-mail: lmahadev@g.harvard.edu}
}
\title{Programming shape using kirigami tessellations}

\date{} 

\baselineskip24pt

\maketitle

\begin{sciabstract}
Kirigami tessellations, regular planar patterns formed by cutting flat, thin sheets, have attracted recent scientific interest for their rich geometries, surprising material properties and promise for technologies. Here we pose and solve the inverse problem of designing the number, size, and orientation of cuts that allows us to convert a closed, compact regular kirigami tessellation of the plane into a deployment that conforms approximately to any prescribed target shape in two and three dimensions.  We do this by first identifying the constraints on the lengths and angles of generalized kirigami tessellations which guarantee that their reconfigured face geometries can be contracted from a non-trivial deployed shape to a novel planar cut pattern. We encode these conditions in a flexible constrained optimization framework which allows us to deform the geometry of periodic kirigami tesselations with three, four, and sixfold symmetry, among others, into generalized kirigami patterns that deploy to a wide variety of prescribed boundary target shapes. Physically fabricated models verify our inverse design approach and allow us to determine the tunable material response of the resulting structures. We then extend our framework to create generalized kirigami patterns that deploy to approximate curved surfaces in $\mathbb{R}^3$. Altogether, this work illustrates a novel framework for designing complex, shape-changing sheets from simple cuts showing the power of kirigami tessellations as flexible mechanical metamaterials.
\end{sciabstract}

Kirigami is the creative art of paper cutting and folding that originated in Japan. Kirigami tessellations, which are regular planar patterns formed by cuts, have recently emerged as paradigms of mechanical metamaterials. Various studies have focused on their geometry and kinematics \cite{Grima00,Grima04,Grima05}, the mechanics of their deployment \cite{Rafsanjani16,Sussman13,Blees15,Zhang15,Shyu15,Rafsanjani17}, the mathematics of their constructions \cite{Konakovic16}, and as materials with a range of unusual properties such as topological insulators \cite{Kane14,Sun12}, auxetic structures \cite{Tang17,Gatt15}, super-stretchable sheets \cite{Kolken17,Mitschke13,Shan15,Isobe16,Neville16} etc. However, almost without exception, these studies study the forward problem of understanding the behavior given the topology and geometry of the tessellations. From both a mathematical and a technological perspective, a very natural question that arises is this: how can one design the geometry of the cuts in a given planar structure, i.e. tessellate a closed, compact subset of the plane, so that it can be deployed into a prescribed final shape in two or three dimensions? Here, we answer this question using an inverse approach couched as a constrained optimization problem, that has some algorithmic similarities to our recent solution of the inverse problem for origami tessellations for programming curvature \cite{Dudte16}.

We start from the well known observation that the only regular polygons that can tile the plane are triangle, square and hexagon. While any of these simple tilings and even more complex ones \cite{Grunbaum87,Rafsanjani16} can be the basis for a deployable kirigami pattern, to simplify our discussion, we present our framework using the quad kirigami pattern (see Section S1 in Supplementary Information - SI for the others). In Fig. \ref{fig:F1}{\bf a}, we show the quadrilateral tessellation of the plane in its compact and deployed state, with the cuts along the edges of the quads designed to allow for rotational in-plane deployment about hinges. The deployment associated with thus 1-degree of freedom mechanism yields a continuous family of self-similar shapes that is geometrically self-limiting eventually. The basic unit cell underlying this pattern is also shown in Fig. \ref{fig:F1}{\bf a} in both the undeployed and deployed states and clearly shows the sides that must be identified together and the hinges about which the quads rotate, both of which provide constraints on the class of admissible solutions.  

%\section*{Inverse planar kirigami design}
Our inverse design problem then becomes this: how should the unit cell be modulated in space to approximate a given planar shape in its deployed state and still be able to tile the plane compactly when undeployed. In Fig. \ref{fig:F1}{\bf b}, we outline this procedure. To find a pattern that can deploy to approximate a prescribed planar shape, it is natural to perform the search for the admissible results in the deployed space. Therefore, we start with some approximation to the given deployed configuration that matches the prescribed boundary curve that might be obtained using, for example, a conformal/quasi-conformal map \cite{Choi18,Meng16} (although we could use any approximation that preserves the number and connectedness of the quads).  In general the deployed configuration  will not correspond to a deployment of a kirigami pattern as the associated length and angle constraints at every interior node are not satisfied. Imposing these interior and boundary conditions leads to a constrained optimization problem whose solution naturally leads to an admissible deployed kirigami pattern whose compact tiling follows via a simple, albeit non-affine contraction. 

%For clarity, we use Greek letters ($\alpha, \beta,...$) for angles, English letters ($a,b,...$) for edge lengths, bold English letters (${\bf p}, {\bf q}, ...$) for the coordinates of nodes, and vectored letters ($\vec{u}, \vec{v}, ...$) for edges and vectors.

%\subsection*{Boundary shape matching constraints}
To ensure that for a given boundary curve $\p S$ we have a valid deployed configuration, we formulate the constraint as the \emph{boundary shape matching} constraints. The constraints force all nodes on the boundary of the deployed configuration to lie exactly on $\p S$. Mathematically, for every boundary node $\mathbf{p}_i$, we should have
\begin{equation}\label{eqt:boundary}
\|\mathbf{p}_i - \widetilde{\mathbf{p}}_i\|^2 = 0,
\end{equation}
where $\widetilde{\mathbf{p}}_i$ is the projection of $\mathbf{p}_i$ onto $\p S$ and $\| \cdot \|$ is the Euclidean 2-norm. In addition to matching the target boundary shape in the deployed configuration, we can also control the boundary shape of the kirigami tessellations in the undeployed state, although here we neglect this additional condition (see Section S2 in SI for details of how to incorporate this as well).

%\subsection*{Contractibility constraints}
For a valid deployed configuration, we must also be able to \emph{contract} (undeploy) the configuration into a  generalized kirigami pattern that closes consistently along the cuts without any mismatch or overlap in lengths and angles. As illustrated in Fig. \ref{fig:F1}{\bf b}, for a valid deployed configuration of a generalized kirigami pattern, the following \emph{contractibility} constraints should be satisfied: (i) for every pair of edges with edge lengths $a, b$ in the deployed space that correspond to the same cut,
\begin{equation}\label{eqt:length}
a^2 - b^2 = 0.
\end{equation}
(ii) For every set of four angles in the deployed space that correspond to an interior node, their sum should be $2\pi$:
\begin{equation}\label{eqt:angle}
\theta_1 + \theta_2 + \theta_3 + \theta_4 = 2\pi,
\end{equation}
where $\theta_i$ are angles in the deployed space as illustrated in Fig. \ref{fig:F1}{\bf a}. The analogous formulation for other kirigami tessellations are discussed in Section S2 in SI.

%\subsection*{Non-overlap constraints}
While the constraints described above ensure a consistency between corresponding edges and angles, they do not prevent the faces from overlapping. To enforce this, we use the following \emph{non-overlap} constraints at every angle between two adjacent faces:
\begin{equation}\label{eqt:nonoverlap}
\langle (\mathbf{b}-\mathbf{a}) \times (\mathbf{c}-\mathbf{a}), \vec{n} \rangle \geq 0,
\end{equation}
where $\mathbf{a}$ and $\mathbf{b}$ are two nodes of a face, $\mathbf{a}$ and $\mathbf{c}$ are two nodes of another face, $(\mathbf{b}, \mathbf{a}, \mathbf{c})$ form a positive (right-hand ordered) angle between the two faces, and $\vec{n} = (0,0,1)$ is the outward unit normal.

%\subsection*{Objective function}
Solving a constrained optimization problem involving all the above constraints is sufficient to guarantee a valid deployed configuration of a generalized kirigami pattern that approximates the prescribed shape. However, the solution is likely to be very rough with large gradients in the shapes of the quads. To get a smooth kirigami tessellation, we therefore minimize the following objective function:
\begin{equation}\label{eqt:objective}
\frac{1}{M} \sum_{i=1}^M \left( \sum_j (\alpha_{i_j} - \beta_{i_j})^2 + \sum_k (a_{i_k} - b_{i_k})^2 \right)
\end{equation}
where $\alpha_{i_j}, \beta_{i_j}$ are a pair of corresponding angles in two adjacent cells and $a_{i_k}, b_{i_k}$ are corresponding edge lengths in two adjacent cells, and $M$ is the total number of pairs of adjacent cells, subject to the constraints \eqref{eqt:boundary},\eqref{eqt:length},\eqref{eqt:angle},\eqref{eqt:nonoverlap}. We solve the problem numerically using MATLAB's built-in optimization routine \texttt{fmincon} (see Section S3 in SI for details). {We note that our optimization problem is underconstrained so that there are multiple admissible deployable kirigami patterns (see Section S2 and S4 in SI for details).} Once we find an admissible deployed kirigami pattern, we can contract this into its compact form by rotating the faces contracting the entire structure. 

\begin{figure}[h!]
\centering
\includegraphics[width=\textwidth]{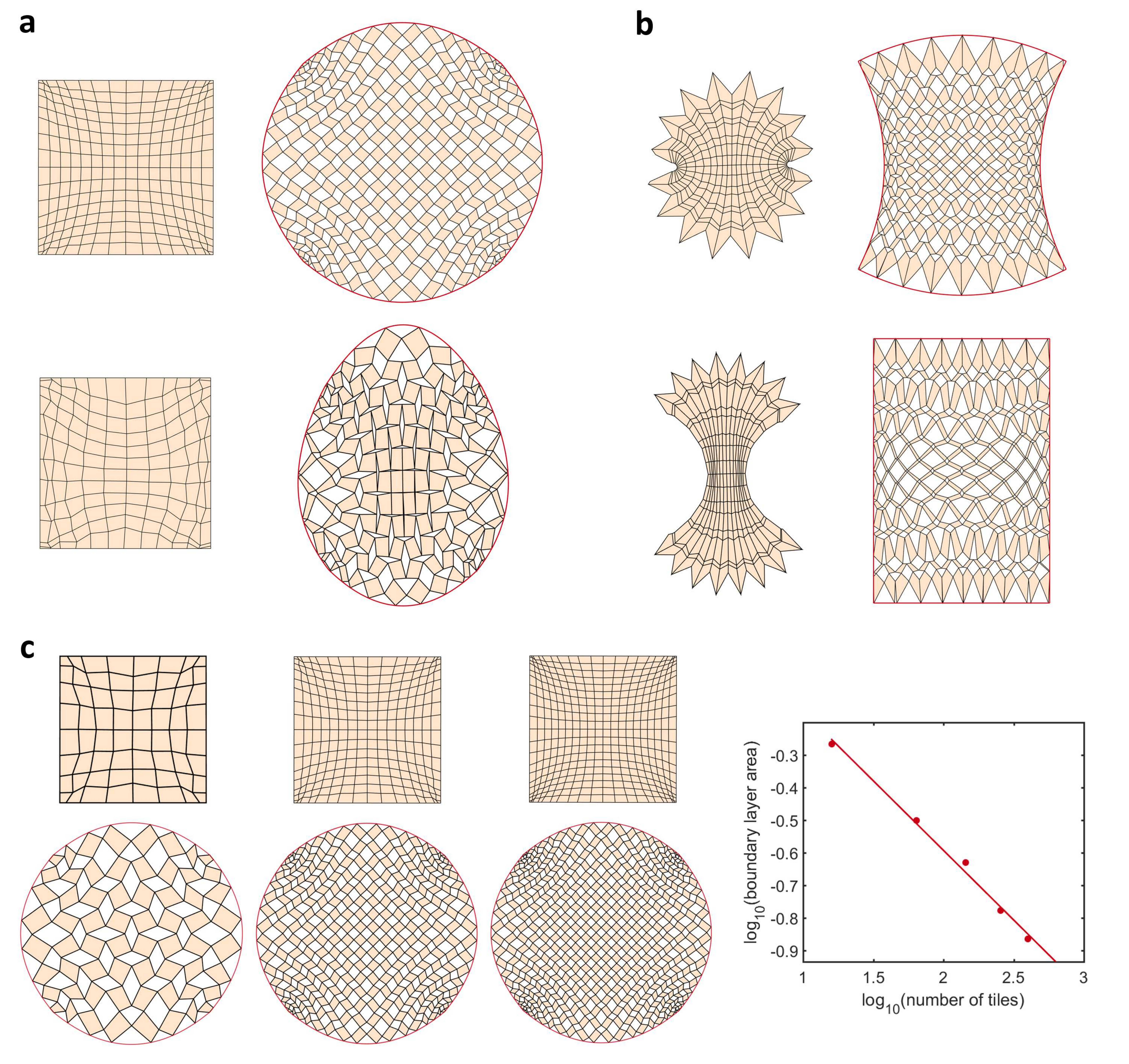}
\caption{{\bf Generalized kirigami patterns.} {\bf a} Examples of generalized kirigami patterns produced by our method for getting a circle or an egg shape from a square upon deployment. {\bf b} Examples of generalized kirigami patterns produced by our method for achieving boundary shapes with mixed curvature or zero curvature. It can be observed that our method is capable of producing generalized kirigami patterns that matches boundary curves with different curvature properties when deployed. {\bf c} Examples of circling the square with different resolutions (number of tiles = $8 \times 8$, $16 \times 16$, $20 \times 20$), together with a log-log plot of the boundary layer area against the number of tiles. Here, the boundary layer area is defined as the total area of the gaps between the circle and the boundary of the deployed kirigami patterns. The dots on the log-log plot represent kirigami patterns with different number of tiles ($4 \times 4$, $8 \times 8$, $12 \times 12$, $16\times 16$, $20\times 20$), and the straight line is the least-square regression line. The result shows that there is an accuracy-effort trade-off in approximating a prescribed shape using generalized kirigami tessellations.}
\label{fig:F2}
\end{figure}

To illustrate the effectiveness of our inverse design approach, we show how to circle the square and how to make an egg shape from a square via generalized kirigami patterns. As shown in Fig. \ref{fig:F2}{\bf a}, by introducing novel generalized kirigami patterns on a square, the deployed configurations can effectively match either a circle or an egg.

Our method can also generate novel generalized kirigami patterns that, when deployed, approximate boundary shapes with different curvatures. Fig. \ref{fig:F2}{\bf b} shows two generalized kirigami patterns; one deploys to approximate a boundary shape with mixed curvature, and the other deploys to approximate a rectangle (see Section S4 in SI for more generalized kirigami patterns with other base tessellations). {These generalized kirigami patterns with different base tessellations exhibit different behaviors in terms of their porosity and magnifying property (see Section S5 in SI for details).}

There is naturally an accuracy-effort trade-off in approximating a prescribed shape using generalized kirigami tessellations: a larger number of smaller tiles allows for improved approximation at the cost of more tiles. Fig. \ref{fig:F2}{\bf c} shows several generalized kirigami patterns of circling the square with different resolution; with more tiles, the boundary of the deployed pattern gets closer to a perfect circle {(see Section S4 in SI for multiresolution results for other patterns)}. To quantitatively assess the accuracy of the patterns for the approximation, we define the boundary layer area of a generalized kirigami pattern by the total area of the gaps between the target boundary shape and the boundary of the deployed configuration. From the log-log plot in Fig. \ref{fig:F2}{\bf c}, it can be observed that as the boundary layer area (denoted by $A$) decreases as the number of tiles (denoted by $n^2$) increases. More specifically, the slope of the least-square line is approximately $-1/2$, which indicates that $A \propto (n^2)^{-1/2} = n^{-1}$. To further explain this, we approximate every boundary gap by a triangle and measure the change in the average triangle base length $\tilde{l}$ and average triangle height $\tilde{h}$ for different resolutions. We observe that $\tilde{l} \propto n^{-1}$ and $\tilde{h} \propto n^{-1}$, and hence the average area of the triangles $\tilde{a} \propto n^{-2}$. As the number of boundary gap triangles is approximately $4n$, we have $A \approx 4n\tilde{a} \propto n^{-1}$. 

\begin{figure}[h!]
\centering
\includegraphics[width=\textwidth]{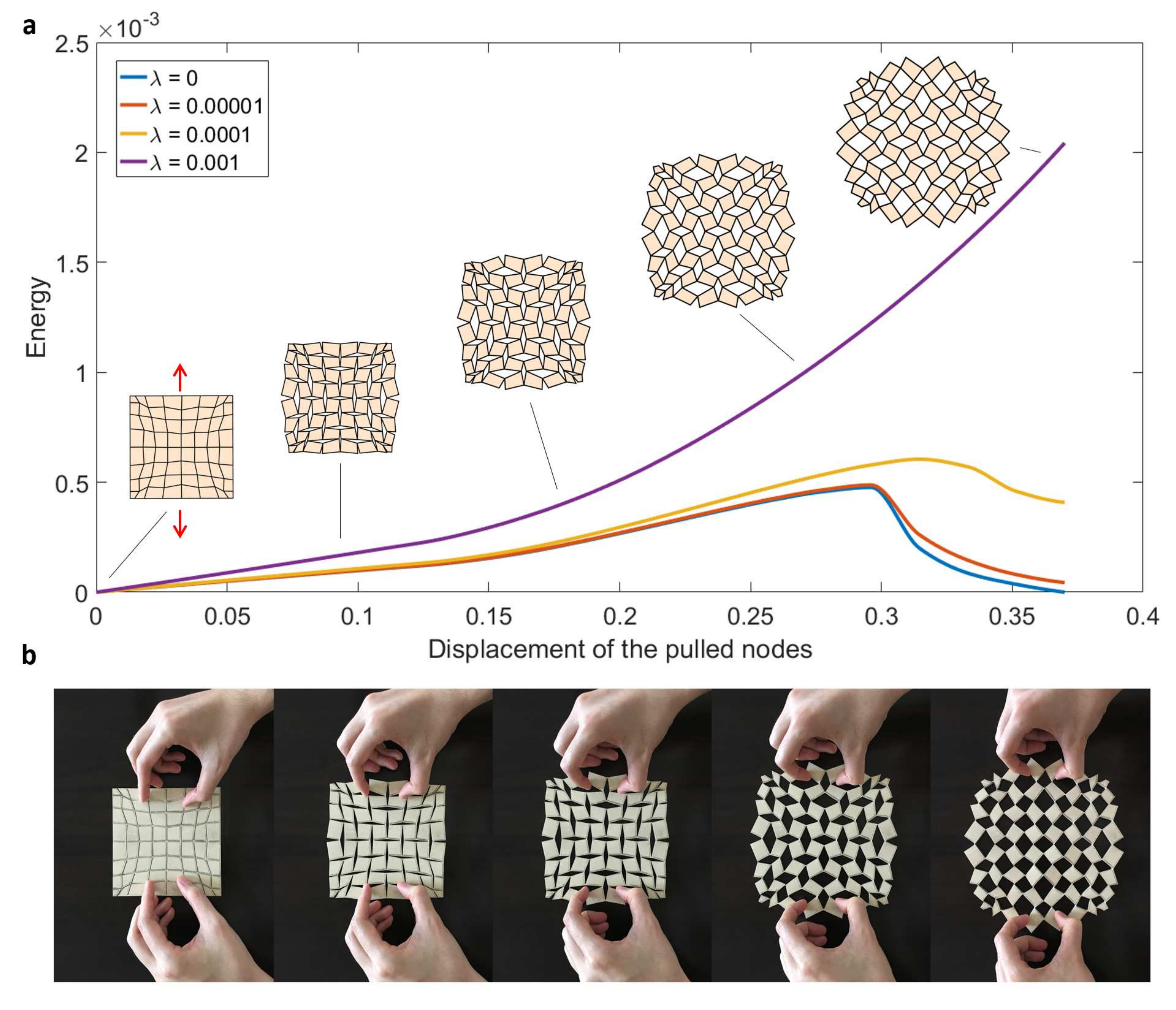}
\caption{{\bf Deployment of generalized kirigami tessellation.} {\bf a} Energetics of the deployment simulations of the square to circle example with different choice of $\lambda$. The insets show the initial, intermediate and final configurations of the generalized krigiami pattern under deployment. {\bf b} Snapshots of the deployment of a monostable fabricated model.}
\label{fig:F3}
\end{figure}

%\section*{Deployment energetic analysis}
Our inverse design approach guarantees that the end-points, i.e. the deployed and undeployed states satisfy the contractibility and boundary constraints, but are agnostic to whether there is a  one-parameter family of zero-energy solutions linking them. To see if this is indeed true, we simulate the deployment process from the initial state to the deployed state by considering a simple physical model of the kirigami tessellation using linear springs along the edges and diagonals of the quads, and simple torsional springs at the nodal hinges to model the ligaments that hold the structure together in real systems.    
% $\mathbf{x}_i$ and $\mathbf{x}_j$, given by $\e_{ij} = \frac{\|\mathbf{x}_i - \mathbf{x}_j\| - l_{ij} }{l_{ij}}$, where $l_{ij}$ is the rest length of the spring. For an intermediate configuration with the perturbed boundary, the strain energy $ E(\mathbf{x}_1, \mathbf{x}_2, ..., \mathbf{x}_N) = \frac{1}{N_s}\sum_{i,j} \left(\frac{\|\mathbf{x}_i - \mathbf{x}_j\| - l_{ij} }{l_{ij}}\right)^2
%$ with $N_s$ being the total number of springs provides us with a measure of the violation of the contractibility constraints.
%
%Note that in real material systems, material connections at hinges have to be taken into account throughout the deployment process. In particular, as cuts tend to close up, there is a tendency for minimizing the angles between cuts. To model this phenomenon, we consider minimizing the following combined energy
Then, the mechanical energy of the system is given by
\begin{equation}
E(\mathbf{x}_1, \mathbf{x}_2, ..., \mathbf{x}_N) =  \frac{1}{N_s}\sum_{i,j} \left(\frac{\|\mathbf{x}_i - \mathbf{x}_j\| - l_{ij} }{l_{ij}}\right)^2 + \lambda \frac{1}{N_c} \sum_{i} \theta_i^2,
\end{equation} 
where $\mathbf{x}_i$ are the locations of the nodes, $\theta_i$ are the angles between every pair of edges created under the cuts, $l_{ij}$ are the rest lengths of the extensional springs, $N_s$ is the total number of extensional springs, $N_c$ is the total number of torsional springs, and $\lambda$ is the ratio of the torsional spring constant to the extensional spring constant. A larger $\lambda$ corresponds to a thicker ligament, which has a larger tendency to close up. By iteratively moving the boundary nodes to the target boundary shape and solving for the intermediate deployed configurations, we obtain a continuous deployment simulation. Fig. \ref{fig:F3}{\bf a} shows the energetics of the deployment simulations with different $\lambda$: if $\lambda \to 0$, we see the presence of bistability, while if $\lambda \not\to 0$, monostability or multistability can be observed. To verify these results, we fabricated a physical model by laser cutting sheets of super-stretchable abrasion-resistant natural rubber. Fig. \ref{fig:F3}{\bf b} shows the deployment snapshots of a fabricated model with monostability, showing that the computational design and the real deployment have similar behaviors (see Section S4 in SI for another example).

%\section*{Inverse 3D kirigami design}
While our inverse design approach was originally focused on approximating planar shapes, it can be extended to fit surfaces in three dimensions, i.e. the pattern space is in $\mathbb{R}^2$ while the deployed space is in $\mathbb{R}^3$.  To fit a surface $S$ in $\mathbb{R}^3$ instead of a given planar boundary shape, we replace the boundary shape matching constraints \eqref{eqt:boundary} by the \emph{surface matching} constraints below. For every node $\mathbf{x}_i$ in the deployed configuration, we enforce
\begin{equation} \label{eqt:surface}
\|\mathbf{x}_i - \widetilde{\mathbf{x}}_i\|^2 = 0,
\end{equation}
where $\widetilde{\mathbf{x}}_i$ is the projection of $\mathbf{x}_i$ onto $S$  and $\| \cdot \|$ is the Euclidean 2-norm. The extra constraints for controlling the boundary shape of the undeployed configuration in the planar case can be directly extended to the surface fitting problem.

One can easily note that the \emph{contractibility} constraints for surface fitting are the same as \eqref{eqt:length} and \eqref{eqt:angle}. For the \emph{non-overlap} constraints, to prevent adjacent faces in the deployed configuration from overlapping or intersecting between each other, we enforce the following inequality constraints for every two adjacent faces in the deployed configuration:
\begin{equation} \label{eqt:nonoverlap3d}
\langle (\mathbf{b}-\mathbf{a}) \times (\mathbf{c}-\mathbf{a}), (\mathbf{c}-\mathbf{a}) \times (\mathbf{d}-\mathbf{a})\rangle \geq 0,
\end{equation}
where $\mathbf{a}, \mathbf{b}$ are two nodes of a face, $\mathbf{c}, \mathbf{a}, \mathbf{d}$ are three nodes of another face, $(\mathbf{b},\mathbf{a},\mathbf{c})$ form a positive (right-hand ordered) angle between the two faces, and $(\mathbf{c},\mathbf{a},\mathbf{d})$ also form a positive (right-hand ordered) angle. The idea is to replace the unit normal $\vec{n}$ in \eqref{eqt:nonoverlap} by the normal computed using one of the two faces.

Note that in the case of deployed configurations for planar shape matching, all faces are automatically planar. This property does not necessarily hold for surface fitting. In addition to the above constraints, we need to enforce the following \emph{planarity} constraints for each face $F$:
\begin{equation} \label{eqt:planarity}
\text{Volume}(F) = 0.
\end{equation}
More explicitly, for quad tessellations the constraint becomes
\begin{equation}
\langle (\mathbf{b}-\mathbf{a}) \times (\mathbf{c}-\mathbf{a}), \mathbf{d}-\mathbf{a}\rangle = 0,
\end{equation}
where $\mathbf{a}, \mathbf{b}, \mathbf{c}, \mathbf{d}$ are the four points of the quad $F$.

Finally, both the objective function \eqref{eqt:objective} for the planar case and the contraction process can be easily extended for surface fitting. Hence, we can obtain a valid generalized kirigami pattern that deploys to approximate a prescribed surface by solving a constrained optimization problem analogous to the planar one using \texttt{fmincon} in MATLAB, with \eqref{eqt:surface},\eqref{eqt:length},\eqref{eqt:angle},\eqref{eqt:nonoverlap3d},\eqref{eqt:planarity} to be satisfied. 

\begin{figure}[h!]
\centering
\includegraphics[width=\textwidth]{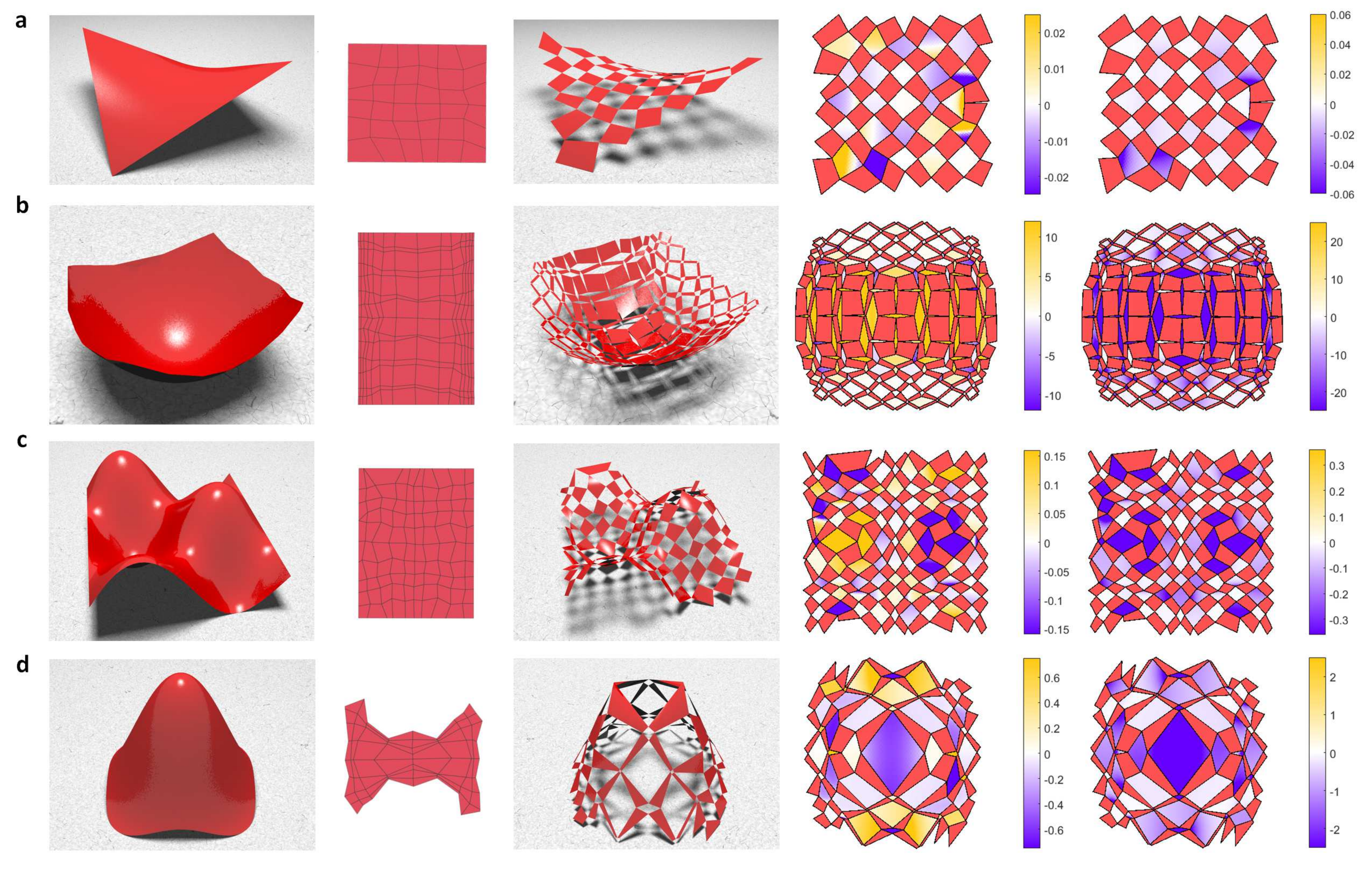}
\caption{{\bf Generalized kirigami patterns for surface fitting.} The target surfaces are {\bf a} a hyperbolic paraboloid (with negative curvature), {\bf b} a paraboloid (with positive curvature), {\bf c} a landscape shape with crests and troughs (with mixed curvature), and {\bf d} a hat-like surface. Columns: The target surfaces (leftmost), the generalized kirigami patterns, the deployed configurations of the patterns that fit the target surfaces, the top views of the deployed patterns with the holes colored with the approximated mean curvature, and the top views of the deployed patterns with the holes colored with the approximated Gaussian curvature (rightmost).}
\label{fig:F4}
\end{figure}

Fig. \ref{fig:F4} shows several generalized kirigami patterns that deploy to fit surfaces with different curvature properties. Again, we can impose extra constraints to further achieve different effects. For instance, we can obtain a rectangular quad pattern that deploys to fit a hyperbolic paraboloid.

It is noteworthy that surfaces in $\mathbb{R}^3$ have non-zero curvature in general, while all faces of the deployed configurations of the generalized kirigami patterns must have zero curvature. This suggests that the curvature of the surfaces resides in the holes of the deployed configurations of generalized kirigami patterns. To verify this, we consider fitting every hole in the deployed configurations of generalized quad kirigami patterns by a bicubic B\'{e}zier surface, and compute the mean curvature and the Gaussian curvature of it (see Section S6 in SI for the details). From the curvature plots in Fig. \ref{fig:F4}, it can be observed that the curvatures of the underlying surfaces indeed reside at the holes of the deployed generalized kirigami patterns, especially at the highly curved regions of the underlying surfaces.

%\section*{Discussion}
Our inverse design approach allows us to create generalized kirigami patterns that approximate a large variety of planar shapes when deployed. Deployment simulations and fabricated models have shown that the generalized kirigami patterns achieve monostability with thick hinges, and by reducing the size of the hinges we can achieve bistability. Our method can also be extended for fitting surfaces in 3D.  By establishing a framework for designing generalized kirigami patterns, our work has set the stage for a better understanding of the interplay between geometry, topology and mechanics.
%
%Also, we have only tackled the 3D surface fitting problem by generalizing the geometric constraints for the planar case without considering the mechanics of the hinges in 3D, which should be important in practice. In addition, we have only focused on patterns without any stochastic variability or noise. All of them are interesting directions to be explored in the future.
%
%The interdisciplinary nature of kirigami has fascinated not only artists but also mathematicians, scientists and engineers.

%We remark that there is a beautiful connection between the quad krigiami pattern and the Miura-ori pattern in origami. The Miura-ori pattern can be generalized to approximate surfaces by imposing suitable constraints on the fold patterns \cite{Dudte16}, and we have shown in this work that the quad kirigami pattern can be generalized to fit planar shapes or surfaces by imposing constraints analogously. Specifically, the angle sum constraints for kirigami patterns are analogous to the developability constraints for Miura-ori patterns. Also, as pointed out by Tachi \cite{Tachi10}, there is in general an obstruction in the folding process of the Miura-ori pattern, which causes bistability. For the generalized kirigami patterns, we have shown in this work that there is also an energy barrier in the deployment process, which leads to bistability in case the connection between faces is close to point connection. In the future, we plan to explore more analogies between the theories of origami and kirigami.

{\bf Acknowledgment} This work was supported in part by the Croucher Foundation (to G.P.T.C.), National Science Foundation Grant DMR 14-20570 (to L.H.D. and L.M.) and DMREF 15-33985 (to L.H.D. and L.M.).

{\bf Competing Interests} We have filed a patent on our algorithms for kirigami design.

{\bf Data Availability} Additional information is available in the Supplementary Information. Correspondence and requests for materials should be addressed to L.M..

\newpage 

\begin{figure}[h]
\centering
\includegraphics[width=\textwidth]{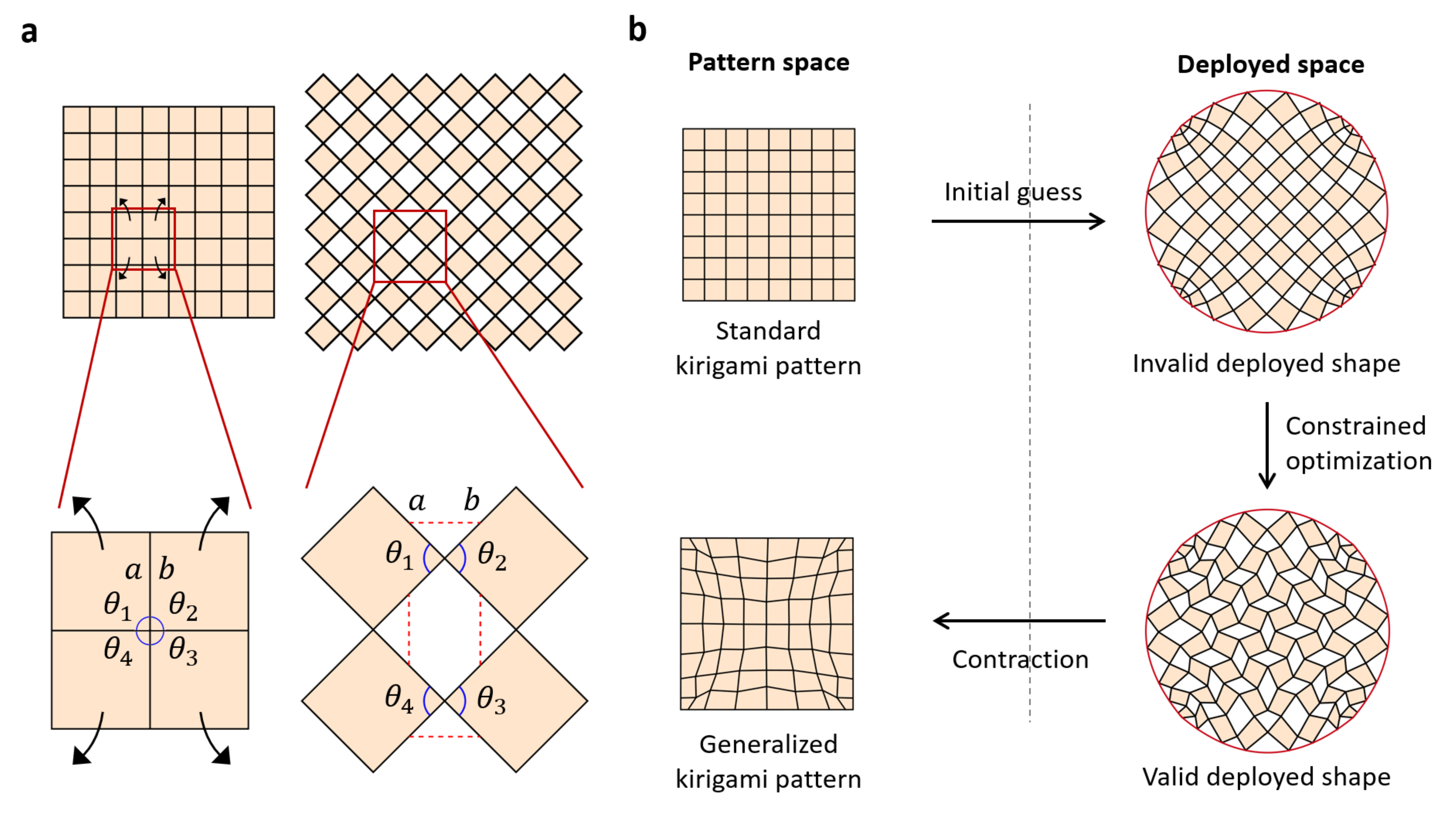}
\caption{{\bf Inverse design framework.} {\bf a} A quad kirigami tessellation and its deployed configuration, with a zoom-in of the unit cell of the quad kirigami tessellation and its deployed configuration. Every pair of corresponding edges are connected by a red dotted line. The set of angles corresponding to the same node are highlighted in blue. In a valid deployed configuration of a generalized kirigami pattern, every pair of edges should be equal in length, i.e. $a = b$, and every set of corresponding angles should add up to $2\pi$, i.e. $\theta_1 + \theta_2 + \theta_3 + \theta_4 = 2\pi$. {\bf b} Our inverse design framework. Given a standard kirigami tessellation, we start with an initial guess in the deployed space. Here the initial guess shown is a conformal map from the standard deployed configuration to the disk. The initial guess is usually invalid, violating either the edge length constraint or the angle constraint, or not exactly matching the target boundary shape. We then solve a constrained optimization problem to morph the initial guess until it becomes a valid deployed shape, satisfying all constraints. Finally, we use a simple contraction procedure to obtain the generalized kirigami pattern.}
\label{fig:F1}
\end{figure}

\end{document}